\documentstyle[12pt,epsf]{article}
\author{A.~A.~Stanislavsky\\
\it
Institute of Radio Astronomy\\
\it of the Ukrainian National Academy of Sciences\\
\it 4 Chervonopraporna St., 61002 Kharkov, {\bf UKRAINE}\\
\it E-mail: alexstan@ira.kharkov.ua}
\title{Memory Effects And Macroscopic Manifestation of Randomness}
\begin{document}
\large\tolerance8000\hbadness10000\emergencystretch3mm
\maketitle

\begin{abstract}
      It is shown that due to memory effects the complex behaviour
of components in a stochastic system can be transmitted to
macroscopic evolution of the system as a whole. Within the Markov
approximation widely using in ordinary statistical mechanics,
memory effects are neglected. As a result, a time-scale separation
between the macroscopic and the microscopic level of description
exists, the macroscopic differential picture is no a consequence
of microscopic non-differentiable dynamics. On the other hand, the
presence of complete memory in a system means that all its
components have the same behaviour.  If the memory function has no
characteristic time scales, the correct description of the
macroscopic evolution of such systems have to be in terms of the
fractional calculus.
\end{abstract}

{\quad\small PACS number(s): 05.40.-a, 05.45.-a, 05.60.-k}

\section{Introduction}
      The connection between microscopic dynamics of components
in stochastic systems and macroscopic description of their
behaviour as a whole is very attractive in statistical
physics\cite{1,2,3}.  From the point of view of a single
trajectory, the path of a Brownian particle is continuous
everywhere, but is nowhere differentiable\cite{4}. Such a path is
not described by an analytical function. In the theory of Brownian
motion formulated by Langevin\cite{5} the velocity of the Brownian
particle was proved to be discontinuous\cite{6}. The
differentiable nature of the macroscopic picture of Brownian
motion is due to the key role of the Central Limit Theorem
(fluctuations of the microscopic quantities are independent of
each other). This means that the microscopic and macroscopic
levels of description of the process are separated in the time
scale, and memory of the non-differentiable character peculiar to
the microscopic dynamics is lost in the long time limit.
Consequently, the results of observing the motion of an ensemble
of trajectories can be predicted by means of theoretical
prescriptions based on ordinary mathematical procedures proceeding
from the differentiability assumption.  When the condition of
time-scale separation is not available, the non-differentiable
nature of the microscopic dynamics can be transmitted to the
macroscopic level\cite{7}.  In the present paper we show that the
key to this understanding is memory effects in stochastic systems.
It is due to memory effects that the macroscopic behaviour of
stochatic systems contains a manifestation of microscopic
dynamics.

     The outline of the paper is as follows. Section~\ref{par2} is
devoted to the behaviour of relaxation in the physical systems
without and with complete memory. In Section~\ref{par3} we use
the generalized (in the terms of the memory effects) Langevin and
Kramers-Moyal equations to shown that, by means of the memory
function, the non-differentiable nature of microscopic dynamics of
system components can be transmitted to the macroscopic level of
description in the form of fractional derivatives. Note that the
memory effects can induce orderedness of macroscopic processes in
stochatic systems.  In Section~\ref{par4} we briefly consider the
criterion of the relative degree of order in the systems.
Section~\ref{par5} is devoted to the Time Fractional
Diffusion-Wave equation concerning the case when chaos and
order coexist. Finally we compare the process described by the
latter equation with the fractional Brownian motion.

\section{Classification of Memory Effects}\label{par2}
      We start from some classification of memory effects. It is
based on mathematical properties of the corresponding memory
function. Let us consider the integro-differential equation
\begin{equation}
\frac{df(t)}{dt}=-\lambda^2\int^t_0K(t-t')\,f(t')\,dt'\,\label{eq1}
\end{equation}
where $f$ is the quantity of interest, $K$ the memory kernel, and
$\lambda$ the parameter. The equation is a typical non-Markovian
equation obtained in studying the physical systems coupled to an
environment, with environmental degrees of freedom being averaged.
The parameter $\lambda$ can be regarded as the strength of the
perturbation induced by the environment of the system.

      For a system without memory (ideal Markov system), the
time dependence of the memory function $K(t-t')$ is of the form
\begin{equation}
K(t-t')=\delta (t-t'),\label{eq2}
\end{equation}
where $\delta(t-t')$ is the Dirac $\delta$-function.  The absence
of memory means that the convolution function,
$J(t)=\int^t_0K(t-t')\,f(t')\,dt'$, is defined by $f(t)$ at the
only instant $t$.  Substituting (\ref{eq2}) into (\ref{eq1}) we
obtain
\begin{equation}
\frac{df(t)}{dt}=-\lambda^2f(t).\label{eq3}
\end{equation}
Equation (\ref{eq3}) has an exponential solution. If memory
effects are introduced into the system the $\delta$-function in
(\ref{eq2}) turns into a bell-shaped function, with the width
determining an interval $\tau$ during which $f(t)$ has an effect
on the function $J$.

      In the literature, wide use is made of the  Markov
approximation, which replaces the equation (\ref{eq1}) with the
rate equation
\begin{equation}
\frac{df(t)}{dt}=-\biggl(\lambda^2\int^\infty_0K(t')\,dt'\biggr)
f(t).\label{eq4}
\end{equation}
The Van Hove limit\cite{8} implies that the limit $\lambda\to 0$,
$t\to\infty$ is approached to in such a way that the product
$\lambda^2t$ is kept constant. Using the Van Hove limit  makes it
possible to replace the time convolution in (\ref{eq1}) with
\begin{equation}
\frac{df(t)}{dt}=-\lambda^2\tau f(t),\label{eq5}
\end{equation} where \begin{displaymath}
\tau=\int^\infty_0dt\,K(t).
\end{displaymath}
The limit $\lambda\to 0$ implies that the coupling of the system
to the environment is weak, while the limit $t\to\infty$ means
that the observation time is much larger than the characteristic
time scale $\tau$.

      On the other hand, in the systems having ideal complete
memory the function $J$ is formed over all the course of
the action of the quantity $f(t')$ up to the instant $t$ with
the weight $K(t-t')=\{1, 0<t'\leq t; 0, t'>t\}$ (step function).
In this case the equation (\ref{eq1}) is transformed into
\begin{equation}
\frac{df(t)}{dt}=-\lambda^2\int^t_0f(t')\,dt'.\label{eq6}
\end{equation}
It has the unique solution, $f(t)=f(0)\cos(\lambda t)$, which
does not decrease at $t\to\infty$ in contrast to (\ref{eq3}).

      Relation (\ref{eq1}) written in the time domain is not
always convenient because of the convolution (integral over
$t'$).  This can be eliminated by using the Laplace transformation
\begin{displaymath}
\bar f(s)={\cal
L}[f(t)]=\int_{0}^{\infty}f(t)\,e^{-st}\,dt,\qquad
f(t)=\frac{1}{2\pi
i}\int^{+i\infty}_{-i\infty}\bar f(s)\,e^{st}\,ds.
\end{displaymath}
In this case the equation (\ref{eq1}) reduces to the
algebraic form
\begin{equation}
s\,\bar f(s)-f(0)=-\lambda^2\bar K(s)\,\bar f(s),\label{eq7}
\end{equation}
where the initial condition is taken into account. The Laplace
transform of the kernel (\ref{eq2}), which corresponds to the
absence of memory, yields the constant, $\bar K(s)=1$.  For ideal
memory, we obtain $\bar K(s)=1/s$.  Thus, as the ideal complete
memory appears in the system the constant kernel is replaced by
the hyperbolic one.  It is logical to infer that the fractional
integration of the order $\nu$, $0<\nu<1$, will interpolate the
memory function between the  $\delta$-function and step function.
Systems with such a memory function occupy an intermediate
position between the two limiting cases and are described in
\cite{9}. They have complete but not ideal memory.  This
means that the memory manifests itself within all the interval
$(0,t)$ but not at each instant $t'$. Such a memory function has no
characteristic time scale, i.\ e.\ the Markov approximation is
inapplicable.

      Assume that memory holds only at the points of a Cantor set.
The problem lies in finding the Laplace transform of a step
memory function. To construct a Cantor set, we first choose
the entire time interval of the length $T$ and remove the central
part of the interval leaving two intervals of the length
$\xi T$ (where $\xi<1/2$).  Obviously, to avoid the loss of the
integral memory, the heights of the two resulting bars must be
increased to the value $(2\xi T)^{-1}$.  In the next stage, each
remaining interval of the length $\xi T$ is subjected to the same
division process. In each subsequent stage $n$, this contraction
procedure is performed for the $2^{n-1}$ stages obtained in the
preceding stage. One can easily see that the memory function
$K_n(t)$ is represented by a set of $2^n$ bars of the height
$1/(2\xi)^nT$ and of the width $\xi^n T$.  The Laplace image of
$K_n(t)$ is written as
\begin{equation}
\bar K_n(s)=\frac{1-\exp(-sT\xi^n)}{sT\xi^n}\,\prod_{k=0}^{n-1}
\frac{1+\exp(-z\xi^k)}{2},\qquad z=(1-\xi)Ts\,.\label{eq8}
\end{equation}
For $n\gg1$ we have $\vert sT\xi^n\vert\ll 1$. As the special
investigations show \cite{9}(Chapter~5), the limiting value of the
function $\bar K_n(s)$, when the number of divisions generating
the Cantor set tends to infinity, becomes
\begin{equation}
\bar K(s)=(sT)^{-\nu}\,q(\ln(sT))\,,\label{eq9}
\end{equation}
where $\nu=\ln 2/\ln(1/\xi)$ is the fractal dimension and
$q(\ln(sT))$ is the periodical function with the period
$\ln\xi$. From the physical point of view the fractal
dimension $\nu$ informs us about the ralative amount of states
being conserved in the process of interaction and represents
a quantitative measure of memory effects. It is clear from the
analysis of the limiting case.  For an empty Cantor set ($\nu=0$)
the dependence (\ref{eq9}) reduces to the constant, corresponding
to the entire absence of memory (in the $t$-domain it consists
of two delta-functions located on the edges of the Cantor set).
The limiting value of the similarity parameter $\xi=1/2$ yields
the dimension $\nu=1$, which corresponds to the complete memory.
The result (\ref{eq9}) is correct for any $sT$ from the interval
\begin{displaymath}
1/(1-\xi)\leq\vert sT\vert<\infty
\end{displaymath}
or for the time variable located in the interval
\begin{displaymath}
0<t/T<1\,.
\end{displaymath}
Averaging the function $q(\ln(sT))$ in (\ref{eq9}) over the period
$\ln\xi$ (see details in \cite{9}) and taking the inverse Laplace
transform of $\bar J(s)$, we obtain the temporal fractional
integral
\begin{displaymath}
J(t)\simeq\frac{1}{\Gamma
(\nu)}\int_0^t (t-t')^{\nu-1}\,f(t')\,dt'\,,
\end{displaymath}
where $\Gamma(\nu)$ is the Gamma function (see Appendix). The
averaging procedure converts the discrete fractal density
$\rho_n=1/(2\xi)^nT$ to its continuous value, $\rho(t)\propto
t^{\nu-1}$. Now the kernel $K(t-t')$ in (\ref{eq1}) becomes
\begin{displaymath}
K(t-t')\sim\frac{1}{\Gamma(\nu)}(t-t')^{\nu-1}
\end{displaymath}
indicating the presence of the fractional integral. The integral
representation of equation (\ref{eq1}) is equivalent to a
differential equation of the fractional order. The above result
(\ref{eq9}) can be generalized to the case of an arbitrary number
$j$ of elementary blocks participating in the construction of a
Cantor set. It turns out\cite{9} that the result (\ref{eq9})
remains true given that $\nu=\ln{j}/\ln(1/\xi)$.  The transition
from the regular Cantor sets to the case, when the parameter $\xi$
is random in each stage of constructing a Cantor set,  leads to
the same result.

      Thus, the systems with residual memory have $\bar
K(s)\simeq s^{-\nu}$, where the exponent value, $0\le \nu\le1$,
determines the extent of memory preservation.  Substituting it
into (\ref{eq7}) (let all constants be contained in $\lambda^2$)
and using the inverse Laplace transformation, the solution of
equation (\ref{eq1}) takes the form:
\begin{displaymath}
f(t)=f(0)\,E_{\nu+1}(-\lambda^2t^{\nu+1})\,,
\end{displaymath}
where
$E_\alpha(z)=\sum_{k=0}^\infty z^k/\Gamma(\alpha k+1)$
is the one-parameter Mittag-Leffler function\cite{10}. In the
particular cases $\nu=0$ and $\nu=1$ we have
\begin{displaymath}
E_1(-z)=e^{-z},\qquad E_2(-z^2)=\cos z.
\end{displaymath}
For $0<\nu<1$ the quantity $f(t)$ has an algebraic decay as
$t\to\infty$. Therefore memory effects can essentially change the
character of relaxation. The above constructions are not only pure
theoretical, but reflect the experimental situation.  In
this connection it should be mentioned that, for example, the
relaxation curves of the experiments\cite{11} on glassy material
with the remarkable accuracy show the algebraic decay rather
than the standard exponential relaxation. The power-law relaxation
can be expected to be a common feature of dynamical systems in a
transition region between the stochastic and regular motion
(supercooled liquids, glasses and polymer materials)\cite{12}.

\section{Qualitative kinetic analysis of random processes with
memory}\label{par3}
      It is well known\cite{13} that the generalized Langevin
equation is of the form
\begin{displaymath}
\frac{dv}{dt}=-\gamma\int^t_0M(t-t')\,v(t')\,dt'+L(t),
\end{displaymath}
where $v$ is the particle's velocity, $M$ some
memory function, $L$ the noise term. If the value $v(t)$ is
observed with a time resolution $\delta t\gg\tau_c$, where
$\tau_c$ is the correlation time of forces producing the random
particle motion, the Markov approximation is applicable, so that
we obtain the ordinary Langevin equation
\begin{displaymath}
\frac{dv}{dt}=-\gamma\,v(t)+L(t).
\end{displaymath}
In the case of complete but not ideal memory the generalized
Langevin equation becomes a fractional differential equation.
The evolution of the probability distribution associated
with the velocity in the fractional Langevin equation is described
by the fractional partial differential equation in the phase space
for the phenomenon\cite{14}.

      Consider the probability distribution $P(x,t)$ to find a
particle at point $x$ and time $t$. The normalization condition is
\begin{equation}
\int\limits_{-\infty}^{\infty}P(x,t)\,dx=1.\label{eq10}
\end{equation}
A broad class of various stochastic processes is given by the
Chapman-Kolmogorov equation:
\begin{equation}
\frac{\partial P(x,t)}{\partial t}=\int M(x,t;x',t')\,
P(x',t')\,d\xi'(t')\,,\label{eq11}
\end{equation}
where the memory function $M(x,t;x',t')$ accounts for the
probability distribution $P(x',t')$ in the previous instantes of
time $t'<t$\cite{15}. The expansion of the kernel $M$ in terms of
the difference $x-x'$ yields the generalized (in the terms of the
memory effects) Kramers-Moyal equation
\begin{equation}
\frac{\partial P(x,t)}{\partial
t}=\sum^\infty_{n=1}(-\nabla)^n \int^t_{-\infty}D^{(n)}(x,t-t')\,
P(x,t')\,dt'\,,\label{eq12}
\end{equation}
where $\nabla\equiv\partial/\partial x$ and the coefficients
$D^{(n)}$ are the moments of the memory function $M$ divided by
$n!$. In the case of Markovian processes the moments are
proportional to $\delta(t-t')$ and the integration in (\ref{eq12})
vanishes. It is relevant to remark that $t$ in (\ref{eq12})
should be treated in the ``kinetic'' but not ``microscopic''
sense.  The Markov approximation (the Van Hove method\cite{8})
ignores memory effects in some sense, but such approach is not
always useful. For example, the statistical theory describing
transport properties of turbulent plasma leads to the conclusion
that the turbulence is of subdiffusive nature and that the
diffusivity considerably decreases.  Therefore, memory effects can
be important for explaining the dependence of the transport
properties of saturated turbulence on the eigenfrequency of the
unstable mode in the case of instability driven by the gradients
in the coordinate space\cite{16}.

      Using (\ref{eq12}) one can write the equation:
\begin{equation}
\frac{\partial P(x,t)}{\partial t}=
\sum\limits_{n=1}^{\infty}(-1)^n\,\frac{1}{n!}\,
\frac{\partial^n}{\partial x^n}\lbrack K_n(x)\,
\int^t_0D(t-t')\,P(x,t')\,dt'\rbrack\,,\label{eq13}
\end{equation}
where $K_n(x)$ are arbitrary functions. The application  of the
Laplace transform to (\ref{eq13}) (with the initial values given
on the whole real axis in the form $P(x,0)$) leads to the
following nonhomogeneous differential equation
\begin{displaymath}
s\bar
P(x,s)-P(x,0)=\bar
D(s)\,\sum\limits_{n=1}^{\infty}(-1)^n\,\frac{1}{n!}\,
\frac{\partial^n}{\partial x^n}\lbrack K_n(x)\,
\bar P(x,s)\rbrack\,.
\end{displaymath}
Consider the case when memory is complete but not ideal (see
previous paragraph). Then we come to the following equation
\begin{equation}
s\bar P(x,s)-P(x,0)=\frac{D_1}{s^\nu}\,
\sum\limits_{n=1}^{\infty}(-1)^n\,\frac{1}{n!}\,
\frac{\partial^n}{\partial x^n}\lbrack K_n(x)\,
\bar P(x,s)\rbrack\,,\label{eq14}
\end{equation}
where $D_1$ is the positive constant. Using the Caputo's
definition (\ref{a11}) of the fractional derivative
(see \cite{17} and Appendix) and its Laplace transform
(\ref{a14}), the study of equation (\ref{eq13}) for the case
of complete but not ideal memory leads to the fractional
generalization of the Kramers-Moyal equation:
\begin{equation}
\frac{\partial^{2\beta}P(x,t)}{\partial t^{2\beta}}=
\sum\limits_{n=1}^{\infty}(-1)^n\,\frac{1}{n!}\,
\frac{\partial^n}{\partial x^n}\lbrack K_n(x)\,
P(x,t)\rbrack.\label{eq15}
\end{equation}
If $K_1(x)$, $K_2(x)$ exist (nonzero) and $K_n(x)=0$ for
$n\geq 3$ the equation (\ref{eq15}) is a fractional generalization
of the Fokker-Planck-Kolmogorov equation\cite{18}.

      Observe some typical features of the fractional
Fokker-Planck-Kolmogorov (FFPK) equation. If $\beta=1/2$ ($\nu=0$)
it is transformed into the convential diffusion equation. This
form is equivalent to the complete absence of memory. For
$\beta=1$ ($\nu=1$) we have the convential wave equation, i.\ e.\
the process with complete memory. The equations containing
derivatives of higher than second order with respect to time
cannot exist in nature: a random process cannot spread faster than
a collection of deterministic trajectories. $\beta=0$ defines the
case of localization, which is the lowest limit of any diffusion
process.  Hence the physical bounds on $\beta$ are given by
$0\leq\beta\leq 1$ (from (\ref{eq14}) it follows that
$1/2\leq\{\beta=(\nu+1)/2\}\leq 1$ because $0\leq\nu\leq 1$).
As is well known from statistical physics, one of simple criteria
of irreversibility is whether or not equations are invariant in
respect to time reversal ($t\to-\,t$).  The specific character of
the processes described by fractional time derivatives is that for
the substitution
\begin{displaymath}
(-t)^{2\beta}=t^{2\beta}\lbrace\cos(2\beta\pi)+
i\sin(2\beta\pi)\rbrace
\end{displaymath}
the relative amount of system states is conserved, and the other
one corresponds to irreversible losses\cite{9}. This allows us to
suppose that for $0<\beta<1$ the FFPK equation describes the
random processes with memory.

\section{Criterion for the relative degree of order}\label{par4}
      The FFPK equation is a integro-differential equation
in partial derivatives with varying coefficients, so in general
one cannot find analytically its solution.  For the  stationary
systems, the probability distribution does not depend on time
($\frac{\partial} {\partial t}P(x,t)=0$), and their analysis leads
to the Gibbs distribution, the cornerstone of statistical physics.
A natural extension of the stationary analysis is the study of
nonstationary systems in the self-similar regime, when the
dependence on two arguments $x,t$ is expressed in terms of a
single variable $y=x/a(t)$:
\begin{equation}
P(x,t)=[a(t)]^\alpha\,\varphi(y)\,,\label{eq16}
\end{equation}
where the functions $a(t),\varphi(y)$ and the exponent $\alpha$ are
to be defined. Mathematically, the probability distribution
(\ref{eq16}) is a homogeneous  function of order $\alpha$.
Physically, the transition to the new variable $y=x/a$ corresponds
to scaling the stochastic quantity $x$ on an arbitrary scale
$a(t)$. It is well known\cite{19} that such a feature is
displayed by fractal objects. If the domain of definition of the
phase space for the stochastic system is a fractal set, its
dimension $D$ lies between 2 (the conventional phase space) and 0
(the point of equilibrium).

      In order to find the exponent $\alpha$, we substitute the
function (\ref{eq16}) into the normalization condition
(\ref{eq10}), getting as a result
\begin{displaymath}
[a(t)]^{-(1+\alpha)}=\int\limits^\infty_{-\infty}\varphi(y)\,dy.
\end{displaymath}
The left-hand side of this equation depends on the time, whereas
the right-hand side does not. Hence, it follows that $\alpha=-1$.
The form of the function $\varphi(y)$ can only be found from the
FFPK equation.

      Of different macroscopic functions, only the entropy $S$
possesses a combination of properties that allow to use it as
a measure of uncertainty (chaos) in the statistical description of
the processes in macroscopic systems\cite{3}.
\begin{displaymath}
S(t)=-\int\limits_{-\infty}^{\infty}P(x,t)\,\log
(P(x,t))\,dx+S_0=
\end{displaymath}
\begin{equation}
=-\int\limits_{-\infty}^{\infty}\varphi(y)\,\log(
\varphi(y))\,dy+\log(a(t))+S_0\,.\label{eq17}
\end{equation}
Define the mean value of $x^2$ as
\begin{displaymath}
\langle x^2\rangle= a^2(t)\int\limits^\infty_{-\infty}
y^2\,\varphi(y)\,dy=B\cdot a^2(t).
\end{displaymath}
Using the above relation we can write the expression (\ref{eq17})
in the form
\begin{equation}
S(t)=-\int\limits_{-\infty}^{\infty}\varphi(y)\,\log(
\varphi(y))\,dy-0.5\log B+0.5\log\langle x^2\rangle+
S_0\,.\label{eq18}
\end{equation}
The entropy $S(t)$ depends on time only on account of $\langle
x^2\rangle$. To consider the evolution of stationary states by
means of slowly varying controlling parameters (they can be found
among the parameters characterizing the stationary state) one can
use the S-theorem \cite{20} as a criterion of the relative
degree of order in various states which reveals for what states the
degree of order is higher. It should be noted that the S-theorem
considers only the stationary states using thermal equilibrium  as
the reference point for the degree of chaos. At $t\to\infty$ the
probability distribution becomes negligible and it is necessary to
renormalize the entropy:
\begin{equation}
\tilde S=S(t)-0.5\log\langle x^2\rangle\,.\label{eq19}
\end{equation}
The  procedure is equivalent to fixing the value $\langle
x^2\rangle$ for any value of a chosen controlling parameter (by
the way, one of conditions of the S-theorem is the equality
condition for the average effective Hamiltonian functions in
different states of interest), since this influences only the
reference point of entropy and does not affect the  renormalized
entropy differences between different states.  Although the value
$\langle x^2\rangle$ depends on the controlling parameter (as well
as on $t$), the fixing of the former does not mean fixing the
latter.  Now using the renormalized entropy difference as a
measure of the relative degree of order, one can study the
evolution of system states in the space of controlling parameters.

\section{Some Examples}\label{par5}
      One of the simple cases of the FFPK equation is the Time
Fractional Diffusion-Wave (TFDW) equation:
\begin{equation}
\frac{\partial^{2\beta}}{\partial
t^{2\beta}}\,P(x,t)= {\it D}\,\frac{\partial^2}{\partial
x^2}\,P(x,t)\,,\label{eq20}
\end{equation}
where ${\it D}$ is a positive constant. Its fundamental solutions
in case of the basic Cauchy and Signalling problems are well
known\cite{21}. Let us take the point $P(\xi,0)=\delta(\xi)$ as an
initial position of a particle. If $1/2<\beta\le 1$, it is
necessary to specify the initial value of the first order time
derivative $\frac{\partial}{\partial t}P(x,t)\vert_{t=0^+}$, since
in this case two linearly independent solutions are to be
determined. To ensure the continuous dependence of our solution on
the parameter $\beta$ in the transition from $2\beta=1^-$ to
$2\beta=1^+$, we assume $\frac{\partial}{\partial
t}P(x,t)\vert_{t=0^+}=0$. Then the equation (\ref{eq20})
has the following solution
\begin{equation}
P(x,t)=\frac{1}{2\sqrt{\it D}t^\beta} \,M\left(\frac{\vert
x\vert}{\sqrt{\it D}t^\beta};\beta\right),\label{eq21}
\end{equation}
where
\begin{displaymath}
M(z;\beta)=\frac{1}{2\pi i}\int_{Ha}e^{\sigma-z\sigma^\beta}
\frac{d\sigma}{\sigma^{1-\beta}},\qquad 0<\beta<1,
\end{displaymath}
$Ha$ denotes the Hankel path (a contour that begins at $\sigma=
-\infty-ia\ (a>0)$, encircles the branch cut that lies along the
negative real axis, and ends up at $\sigma=-\infty+ib\ (b>0)$).

      Mainardi\cite{21} developed the function $M(z;\beta)$
as a series
\begin{displaymath}
M(z;\beta)=\sum^\infty_{n=0}\frac{(-1)^n\,z^n}{n!\,\Gamma
[-\beta n+(1-\beta)]}
\end{displaymath}
and showed that it is a particular case of the Wright
function
\begin{displaymath}
W(z;\lambda,\mu)=\sum^\infty_{n=0}\frac{z^n}{n!\,\Gamma
(\lambda n+\mu)}=\frac{1}{2\pi
i}\int_{Ha}e^{\sigma+z\sigma^\lambda} \frac{d\sigma}{\sigma^\mu},
\end{displaymath}
where $\lambda>-1,\ \mu>0$\cite{10}. It is non-negative for any
$0<\beta\leq 1$ and satisfies the normalization condition
$\int^\infty_0M(\zeta; \beta)\,d\zeta=1$.  Clearly, these
properties are also characteristic to the probability distribution
mentioned above.  For $\beta=1$ the function is the Dirac
$\delta$-function.  For $\beta=1/2$ we have the Gaussian function
$M(z;1/2)=\exp (-z^2/4)/\sqrt{\pi}$. For $0<\beta\leq 1/2$ the
function $M(z;\beta)$ ($z\geq 0$) decreases monotonically, while
for $1/2 <\beta <1$ it first increases and then decreases,
exhibiting its maximum value at a certain point $z_{max}$.
The expression (\ref{eq21}) describes the particle evolution in
the space-time and is the Green function of the TFDW equation.
Averaging $x^2$ we obtain
\begin{equation}
\langle x^2\rangle=\int\limits_{-\infty}^{\infty}
x^2\,P(x,t)\,d\xi=\frac{2\,{\it D}\,t^{2\beta}}
{\Gamma(2\beta+1)}\,.\label{eq22}
\end{equation}
The probability distribution (\ref{eq21}) obeys the
following scaling relations \begin{displaymath} P(\hat x=b^\beta
x,\hat t=bt)=b^{-\beta}\,P(x,t) \end{displaymath} for an arbitrary
parameter $b$.

      Note that if in (\ref{eq21}) $\beta=1/2$ we obtain the
convential Brownian motion (purely random process). For $\beta=1$
the particle performs ballistic motion (purely deterministic
process).  Next we are going to show that $0<\beta<1$ is the
special  case when chaos and ordered motion coexist.

\begin{figure}
\def\epsfsize#1#2{0.6#1}
\centerline{\epsfbox{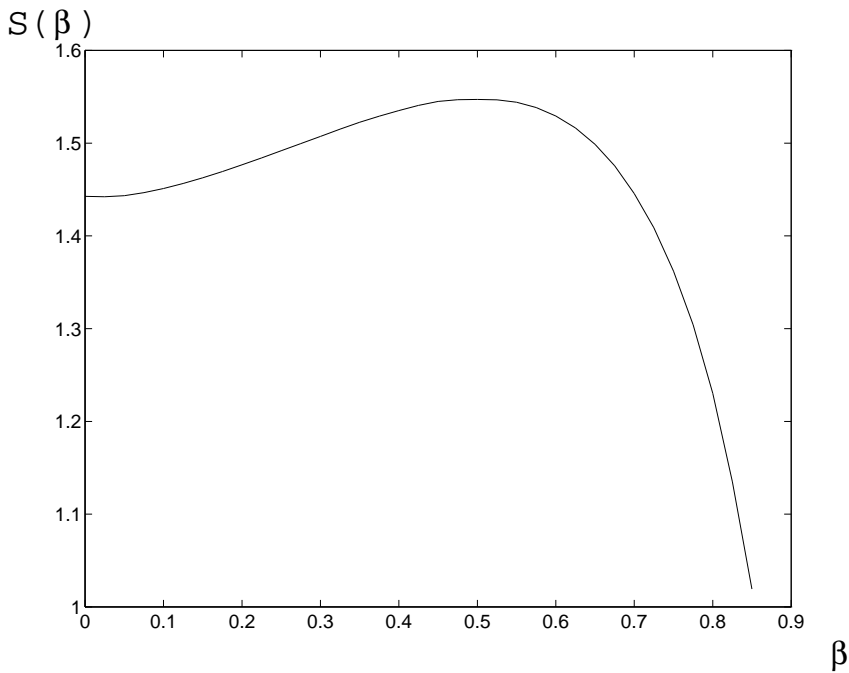}}
\caption[]{The dependence of the value $S(\beta)$ on the
controlling parameter $\beta$.}
\label{fig1}
\end{figure}

      Now let us calculate the entropy (\ref{eq18}) in this case:
\begin{equation}
S(t)=0.5+0.5\log_2(\langle x^2\rangle)+S(\beta)\,.
\label{eq23}
\end{equation}
As a controlling parameter we choose $\beta$ and will consider
the evolution of the sequence of states corresponding to different
values of the controlling parameter. One should keep in mind that
$\lim_{t\to\infty}\langle\xi^2\rangle=\infty$. For this reason we
must renormalize the expression (\ref{eq23}) according to
(\ref{eq19}). Then the relative degree of order can be estimated
as
\begin{equation}
S(\beta)=0.5\log_2\Gamma(2\beta+1)-\int\limits_{0}^{\infty}M(z,
\beta)\,\log_2(M(z,\beta))\,dz.\label{eq24}
\end{equation}
The dependence $S(\beta)$ is represented on Fig.~\ref{fig1}.  It
reaches the maximum at $\beta_c=1/2$. We regard this state as a
state of physical chaos (the correctness of this assumption will
have to be verified). The smooth decrease of the value
(\ref{eq24}) is a quantitative measure of the increase in the
degree of order. Since the inequality $S(1/2)>S(\beta\not=1/2)$ is
satisfied, $\beta\to\frac{1}{2} +\Delta\beta$ is the
transition from a less ordered state (physical chaos) to a more
ordered state. This is an indication that we have found the
corresponding controlling parameter, and the evolution of
the system in the space of the controlling parameter is
associated with self-organization.  The conclusion is valid
because for $\beta=1$ we have the purely deterministic state which
may be taken by a reference point of the degree of order. Some
orderedness also happens to be the case for $0<\beta <1/2$.
Although here we do not come to complete order, in some sense
the orderedness is higher than for $\beta_c=1/2$.  Thus the value
$\beta$ can be adopted as a measure of relative degree of order.
It is useful to mention some particular values, $\lim_{\beta\to
0}S(\beta)=1/\ln 2$ and $S(1/2)=0.5(1+\ln\pi )/\ln 2$. To sum up,
the process considered above is the special case when chaos and
ordered motion coexist, the value $\beta$ characterizing the
relative degree of order in the process. Next we will compare this
process with the fractional Brownian motion.

       In many physical systems the Gaussian distribution is a
straightforward consequence of the Central Limit Theorem which
makes it possible to consider completely random processes. Hurst
found a new set of statistical tools to examine the data which
does not represent a purely random structure though standard
statistical methods do not show any correlation between the
observations\cite{22}.  Using his rescaled range analysis one
can extract meaningful information about the ``memory'' of a
time-series.  If the observations are not independent, each
carries a ``memory'' of events which preceded it. Mandelbrot and
Van Ness\cite{23} suggested the so-called fractional Brownian
motion (the fractional integral of the white Gaussian noise) as a
model reflecting the phenomenon. The fractional Brownian motion
has the following probability distribution
\begin{equation}
\Pi(x;t)=\frac{1}{\sqrt{2\pi\,K}\,t^\beta}\exp\Biggl(
-\frac{x^2}{4K\,t^{2\beta}}\Biggr)\,.\label{eq25}
\end{equation}
Its average $\langle x^2\rangle$ is proportional to
$t^{2\beta}$ as in the case of the process (\ref{eq21}). Then the
expression (\ref{eq18}) is of the form
\begin{equation}
S(t)=0.5\{\ln(2\pi\bar D(t))+1\}\,,\label{eq26}
\end{equation}
where $\bar  D(t)=2K\,t^{2\beta}$. Renormalizing it in accordance
with (\ref{eq19}) we obtain $\tilde S=0.5(\ln2\pi+1)$. In other
words, the renormalized entropy does not depend on $\beta$, i.\
e.\ the fractional Brownian motions with various Hurst exponents
are almost the same in respect to the relative degree of chaos.
This example differs radically from the foregoing one: in the
model of fractional Brownian motion (\ref{eq25}) the rise of
orderedness with changing the Hurst exponent is impossible for
the simple reason that the  shape of the probability distribution
does not change in the space of the controlling parameter (compare
with the process (\ref{eq21})).  The numerical modelling verifies
this conclusion\cite{24}.

\section{Concluding remarks}\label{par6}
       The fractional calculus formalism generalizing
differentiation and integration to fractional orders has a long
history\cite{25}, but recently the interpretative approach opened
broad perspectives in physical and engineering
applications\cite{12,14,18}.  The above consideration shows that
fractional calculus provides a bridge between purely
deterministic processes and purely stochastic ones.  The fact is
of interest in its own right because chaos and order in Nature
coexist.  According to L.  Boltzmann and J.W. Gibbs, in closed
statistical systems, evolution in time results in the equilibrium
state which is the most chaotic (or, in another words, purely
random). The element motions in closed systems are independent of
each other.  In open systems the environment induces memory
effects, so that the macroscopic behaviour of such systems contains
a manifestation of microscopic dynamics. If microscopic motion in
the systems is very complex or random, the complete memory
transmits the complex behaviour to macroscopic evolution of the
system as a whole. It is worth noticing that the model (used by
us) of memory at the points of a Cantor set is not too exotic.  It
implies that memory is intrinsic to all time scales (in such a way
that the corresponding memory function has no characteristic
scales) of the phase space of a system given that the number
of divisions generating a Cantor set tends to infinity.
In a different way, the largest of time scales would be in the
system as in the case of the Markov approximation. We have
demonstrated that the relationship between the Cantor set and the
fractional integral reduces the generalized (in the terms of the
memory effects) Langevin, Kramers-Moyal, Fokker-Planck-Kolmogorov
equations to their fractional form. The fractional generalizations
turn out to be useful for studying the random processes with
residual memory (without any characteristic time scale).

      The process described by the Time Fractional Diffusion-Wave
equation is an example clearly showing that for chaos and ordered
motion in a system to coexist, its probability distribution must
undergo qualitative changes in the function form with slowly
varying the system parameters. In our consideration the parameter
was the relative amount of the system states having orderedness
during the system evolution. The evolution in the space of the
parameter looks like a stochastic analog of bifurcation connected
with the phase transition ``order-disorder''. For $0<\beta<1/2$
the probability distribution has one sharp maximum at $z_{max}=0$.
If the parameter attains the value $\beta=1/2$ the function
becomes flat.  Then for $1/2<\beta<1$ the probability function
takes the form with two maxima symmetrical in respect to the
origin of coordinates (where the function minimum is found).  When
the parameter $\beta$ goes towards 1, the probability to find a
particle between the maxima is becoming less and lesser, and
the function peaks get narrower and higher. For $\beta=1$ the
probability distribution is transformed in two $\delta$-functions.
As a result, this system state becomes completely ordered. The
fractional Brownian motion has not the possibility. The Hurst
exponent (which is the only parameter of the process) influences
the asymtotic behaviour of its autocorrelation function, but the
process remains Gaussian anyhow.

\section*{Acknowledgments}
Portions of this research were reported at the XXVI General
Assembly of URSI (14-21 August 1999, Toronto, Canada). I
appreciate helpful discussions with Prof. Raoul Nigmatullin
(Jordan State University) and thank Prof. Francesco Mainardi
(University of Bologna) for useful comments.

\begin{appendix}
\begin{center} \bf APPENDIX \end{center}
    In this appendix we consider briefly the basic formulas
used in the fractional calculus introduced in the text. The
starting  point is the introduction of the causal function
$\Phi_{\lambda}(t)$ defined as
\begin{displaymath}
\Phi_\lambda(t):=\cases{\frac{t^{\lambda-1}}{\Gamma(\lambda)},&$t\ge 0$;\cr
0,&$t<0$;\cr}\qquad\lambda\in{\bf C},
\end{displaymath}
whose Laplace transform is
\begin{equation}
{\cal L}[\Phi_\lambda(t)]=\bar\Phi_\lambda(t):=
\int^{+\infty}_{0}e^{-st}\,\Phi(t)\,dt=\frac{1}{s^\lambda},
\quad {\rm Re}\,\lambda>0,\quad {\rm Re}\,s>0.\label{a2}
\end{equation}
The function satisfies the composition rule
\begin{displaymath}
\Phi_\lambda(t)*\Phi_\mu(t)=\int^t_0\Phi_\lambda(\tau)\,
\Phi_\mu(t-\tau)\,d\tau=\Phi_{\lambda+\mu}(t),\quad
{\rm Re}\,\lambda>0,\quad {\rm Re}\,\mu>0.
\end{displaymath}

    The integral of order $n$ of a causal function $f(t)$ can be
expressed by the convolution between $\Phi_n$ and $f$
\begin{eqnarray}
I^nf(t)&=&\int^t_0dt_1\int^{t_1}_0dt_2\cdots\int^{t_{n-2}}_0dt_{n-1}
\int^{t_{n-1}}_0f(\tau)\,d\tau=\nonumber\\
 &=&\frac{1}{(n-1)!}\int^t_0f(\tau)\,
(t-\tau)^{n-1}\,d\tau=\Phi_n(t)*f(t),\nonumber
\end{eqnarray}
basing on the well known formula that reduces the calculation of
the $n$-fold primitive to a single integral. For $\lambda>0$, the
function $\Phi_\lambda(t)$ is locally absolutely integrable in
$0\leq t<+\infty$. To extend the above formula from positive
integer values of the index to any positive real values, let us
define the fractional integral of order $\alpha>0$:
\begin{equation}
I^\alpha f(t):=\frac{1}{\Gamma(\alpha)}\int^t_0(t-\tau)^{\alpha-1}
\,f(\tau)\,d\tau=\Phi_\alpha(t)*f(t). \label{a5}
\end{equation}
The Laplace transform of the fractional integral is the
straigthforward generalization of the ordinary case $\alpha=n$
\begin{equation}
{\cal L}[I^\alpha f(t)]=\frac{\bar f(s)}{s^\alpha}. \label{a6}
\end{equation}

      For $\lambda\leq 0$, the causal function $\Phi_\lambda(t)$
is not locally absolutely integrable and consequently the integral
\begin{equation}
\frac{1}{\Gamma(-\alpha)}\int^t_0\frac{f(\tau)}{(t-\tau)^{1+\alpha}}
\,d\tau=\Phi_{-\alpha}(t)*f(t),\quad\alpha\in{\bf R^+}\label{a7}
\end{equation}
is in general devergent. Nevertheless, if $\lambda=-n,\
(n=0,1,\dots)$, the functions $\Phi_\lambda(t)$ can treat in the
framework of generalized functions\cite{26}. In this case it
reduces to the $n$-derivative (in the generalized sence) of Dirac
$\delta$-function
\begin{displaymath}
\Phi_{-n}(t):=\delta^{(n)}(t).
\end{displaymath}
Then formally the derivative of order $n$ of a causal function $f(t)$
can be obtained by the convolution between $\Phi_{-n}(t)$ and $f$
\begin{displaymath}
\frac{d^n}{dt^n}f(t):=\int^t_0f(\tau)\,\delta^{(n)}(t-\tau)\,d\tau=
\Phi_{-n}(t)*f(t),\quad t>0.
\end{displaymath}
The limit case $\alpha=0$ defines the Identity operator
\begin{displaymath}
\int^t_0f(\tau)\,\delta(t-\tau)\,d\tau=
\Phi_0(t)*f(t)=f(t).
\end{displaymath}

    In order to obtain a definition for the fractional
derivative (with non-integer $\alpha$) that is valid for classical
functions, we have to regularize the divergent integral (\ref{a7})
in some way.  As a consequence we arrive at two alternative
regular definitions for the fractional derivative $D^{\alpha}$,
which read for $n-1<\alpha<n$
\begin{eqnarray}
D_c^{\alpha}f(t)&=&\frac{1}{\Gamma(n-\alpha)}\frac{d^n}{dt^n}
\int^t_0\frac{f(\tau)}{(t-\tau)^{\alpha+1-n}}\,d\tau,\label{a10}\\
D_r^{\alpha}f(t)&=&\frac{d^\alpha}{dt^\alpha}f(t)=\frac{1}
{\Gamma(n-\alpha)}\int^t_0\frac{f^{(n)}(\tau)}
{(t-\tau)^{\alpha+1-n}}\,d\tau.\label{a11}
\end{eqnarray}
The difference between the two definitions lies in
\begin{displaymath}
D_c^{\alpha}f(t)=D_r^{\alpha}f(t)+\sum_{k=0}^{n-1}f^{(k)}(0^+)\,
\Phi_{(k-\alpha+1)}(t).
\end{displaymath}
The defintion (\ref{a10}) is the most commonly adopted in
mathematically oriented papers and books\cite{26}. The definition
(\ref{a11}), introduced by Caputo\cite{17}, is more restrictive
than (\ref{a10}) because it requires the function $f(t)$ to be
n-defferentiable. As it was shown in this paper, the latter
definitions is more suitable for the problems considered above
where the conventional initial conditions are expressed in terms
of integer derivatives. Using the classical technique of Laplace
transform
\begin{equation}
{\cal L}\Bigr[\frac{d^n}{dx^n}f(t)\Bigl]=s^n\bar
f(s)-\sum_{n=0}^{n-1} s^{n-1-k}f^{(k)}(0^+),\quad n\in{\bf
N}\label{a13}
\end{equation}
and (\ref{a2}), (\ref{a11}) we get
\begin{equation}
{\cal L}\Bigr[\frac{d^\alpha}{dt^\alpha}f(t)\Bigl]=s^\alpha\bar
f(s)- \sum_{n=0}^{n-1}s^{\alpha-1-k}f^{(k)}(0^+),\quad
n-1<\alpha<n, \label{a14}
\end{equation}
first stated by Caputo\cite{17}. It is worth noticing that
according the definition (\ref{a10}) the fractional derivative of
a constant does not vanish if $\alpha$ is not integer, while
according to the defintion (\ref{a11}) it vanish for any $\alpha$.
\end{appendix}

\end{document}